# Digital LDO with Time-Interleaved Comparators for Fast Response and Low Ripple

Sohail Ahasan[1], Saurav Maji[1], Kaushik Roy[2], Mrigank Sharad[1]
[1]Department of Electronics and ECE, IIT Kharagpur, West Bengal, India
[2]School of Electrical Engineering, Purdue University, West Lafayette, IN, USA

*Abstract*— On-chip voltage regulation using distributed Digital Low Drop Out (LDO) voltage regulators has been identified as a promising technique for efficient power-management for emerging multi-core processors. Digital LDOs (DLDO) can offer low voltage operation, faster transient response, and higher current efficiency. Response time as well as output voltage ripple can be reduced by increasing the speed of the dynamic comparators. However, the comparator offset steeply increases for high clock frequencies, thereby leading to enhanced variations in output voltage. In this work we explore the design of digital LDOs with multiple dynamic comparators that can overcome this bottleneck. In the proposed topology, we apply time-interleaved comparators with the same voltage threshold and uniform current step in order to accomplish the aforementioned features. Simulation based analysis shows that the DLDO with time-interleaved comparators can achieve better overall performance than variable step algorithm based DLDO in terms of current efficiency, ripple and settling time. For a load step of 50mA, a DLDO with 8 time-interleaved comparators could achieve an output ripple of less than 5mV, while achieving a settling time of less than 0.5us. Load current dependant dynamic adjustment of clock frequency is proposed to maintain high current efficiency of ~97%.

*keywords— low power, voltage regulation, VLSI, digital, circuit design*

## I. INTRODUCTION

To meet the demand of energy-efficient computing with scaling CMOS technology, need for distribution and regulation of relatively low on-chip supply voltage has became crucial [1, 2]. Specifically, emerging demand for low power, near threshold computing has posed several challenges of regulation of these voltages (e.g., 320 mV ± 50mV) for highly integrated computing systems like Chip Multi Processors (CMP) [3]. The need for efficient voltage regulation is more pronounced in near sub-threshold computing regime where the gate delay is highly susceptible to supply variation. An increasing number of power domains and of power states per domain, as well as decreasing decoupling capacitance per local grid and wide range of digital load currents necessitate the design of high-efficiency, compact on-die voltage regulators [5,6]. The on-chip LDO regulators are more suitable for the near-threshold/sub-threshold logic circuits [7], since they can supply more stable and precise voltage with lower voltage ripple and faster transient response despite lower power efficiency, compared with the switching regulators [8], [9].Conventional Analog LDOs are not applicable at such low voltages mainly because of increase of PVT variations, poor noise characteristics and the small bias current, mainly in the sub-threshold regime [10],[11].

Digital LDO topologies have been explored in recent years, that can be suitable for low operating voltages([7], [8]). The rationale behind such designs is to convert the control section of an LDO into a compact and scalable digitally implementable circuit ([9]-[11]). The supply devices acting as linear region ON-OFF switches, can operate with lower drop-out voltage, leading to higher efficiency. Dynamic comparators employed in DLDOs can operate at a faster rate, while burning relatively small static power, thereby providing appreciable current efficiency. Further, relatively compact and robust design of digital control units allows the designer to replicate and distribute such regulators in larger numbers on the die to provide ultra-fine grained spatio-temporal power management. It is observed that faster operation of comparators can achieve improved ripple and quicker response to load transients. However, at higher clock frequencies, the power consumption and offset of the dynamic comparators increases steeply, which can degrade the LDO efficiency and increase the output ripple.

To address the above bottlenecks, in this paper we propose time-interleaved LDO model with dynamic frequency adjustment based on Load current level. In section-II we will describe the basic architecture of variable step algorithm based LDO, along with performance analysis of the proposed design. In section-III we will describe the basic architecture of time interleaved LDO, along with performance analysis of the proposed design. In section III system level simulation results are presented. Section IV concludes the paper.

## II. VARIABLE SHIFT ALGORITHM BASED LDO:

### A. DLDO circuit with variable shift control scheme:

This proposed model comprises of a set of driver PMOS transistors (256 used in the simulations) that are directly attached to the point whose voltage is desired to be regulated. The model comprises of a fixed number of Digital Latched Comparators (shown in Fig.1) with the reference voltages equally spaced and encompassing a range centered about the fixed voltage which we desire to achieve (e.g. when the desired voltage is 800 mV,we have a set of 17 comparators,one having reference voltage = 800 mV, 16 comparators in the range 760-840mV spaced equally).

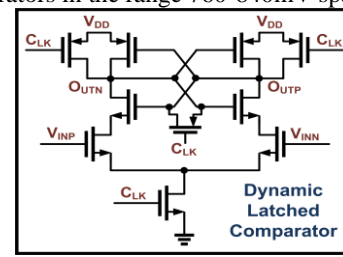

Fig: 1 Dynamic Latched Comparator used in the proposed Models

The operation of the circuit and its control lies in the fact that variable number of comparators can be turned on and off depending on the deviation from the desired voltage and hence is expected to give faster settling time. This is accomplished by a digital logic circuit that has also been simulated in Synopsys. It comprises of a logic that identifies the junction in the thermometer code of the set of PMOS drivers(ie if on is represented by 1 and off is represented by 0, the state of the PMOS drivers comprises of a set of 1s followed by a set of 0s).After identifying the junction and on the basis of the voltage sampled in the present clock cycle the circuit changes the state of some transistors(either turning some on or some off and shifting the thermometer code).This change of the state is variable unlike the Conventional LDO and is dependent on the deviation from the desired voltage. The maximum number of PMOS transistors that can be turned on/off (called the Max Step Size) is fixed by the Digital logic used. Although a wide range for the Maximum step size can be selected, the minimum step size is dependent on the Max Step Size and the number of Comparators. However it can be set to 1(for obtaining high accuracy) by adjusting these parameters.

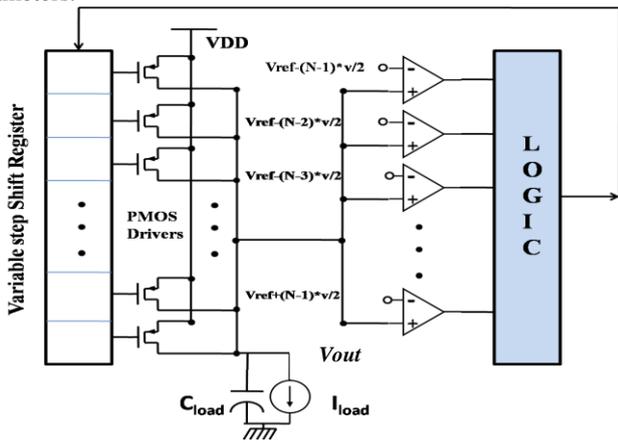

Fig: 2 Schematic Diagram of the ProposedVariable Shift LDO Model

The operation of this Model is described as below. At every clock cycle, the voltage of the desired point is sampled which is fed back to all the comparators .Depending on the Reference voltages of the comparators, some will give VDD or GND as the Output.(e.g. if the voltage sampled is 795mV, Comparators having Vref above 795mV will give GND as the output while comparators below 795mV will give VDD as Output). Now if the voltage is below Vdesired, the more number of comparators will have GND as Output and conversely if the voltage is above Vdesired, the less number of comparators will have GND as Output. The number of PMOS that will be turned on/off depends on the deviation of the states of the comparators from the ideal configuration. Thus depending on the states of the comparators as deviated from the desired configuration the Digital logic circuit would connect its output to the PMOS drivers near the thermometer code junction point. This is because if the Vout is less than Vdesired , more PMOS transistors must be turned on to supply more current and raise the voltage and conversely if the Vout is more than Vdesired, some PMOS transistors must be turned off to reduce the supplied current and lower the voltage. In our proposed model, the digital logic have been designed such that if Vout is greater than Vdesired, the number of PMOS to be turned off = Max Step Size*(Extra number of Comparators that have GND as output/No of Comparators on either side of Vdesired).

Similarly if Vout is less than Vdesired, the number of PMOS to be turned on = Max Step Size*(Extra number of Comparators that have VDD as output/No of Comparators on either side of Vdesired). This proportional change to the deviation ensures proper control. For example, if the voltage have risen to level above the highest Vref, Max Step Size(also called Block Size)number of PMOS transistors will be turned off in the present clock cycle.

Thus depending on the voltage reached by this transition, appropriate action will be taken in the next clock cycle (Time period = 1ns used in this case).The Digital Logic connects the output of the Comparators to the appropriate number of PMOS transistors.

In this proposed scheme the Min Step size(ie minimum no of transistors that can be turned on/off = Max_Step Size*1/(Number of Comparators on either side of Central Comparator).However using more refined Digital logic Min Step Size can be set to 1 for any Max Step Size or number of Comparators but that will not give proportional change in response. For further discussion, we would refer Num-C as the number of Comparators on either side of Central comparators,Vref-Gap as the difference in reference voltage gap between the successive Comparators and Block-Size as the Maximum Step size. Thus Total Number of Comparators = 2*Num-C + 1

*B. Variation with design parameters:*

The performance of the circuit can be measured by the Voltage ripple at the steady state and its settling time(time to attain steady state from zero voltage).A number of design parameters should be properly designed to optimize the performance of the circuit.

*1)Number Of Comparators:*

More number of comparators pave the way for more fine grained voltage reference levels and more fine grained step currents through PMOS switches. Therefore for a constant load (I=10mA, C=5.0 nF) ripple voltage decreases and settling time increases with increase in number of comparators.

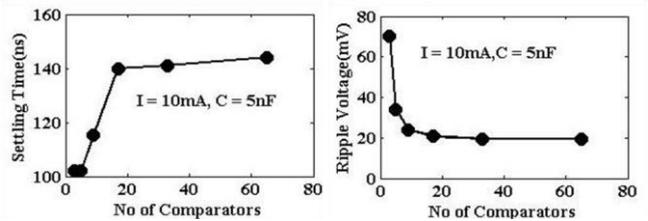

Fig: 3 Variation of settling Time and Ripple Voltage with No of Comparators

*2)Variation with Capacitance:*

For a constant supply of current the charging and discharging times increase proportionately with increase in capacitance attributing for an increase in settling time. Voltage ripple also decreases as higher capacitance filter out the high frequency components of steady state voltage.

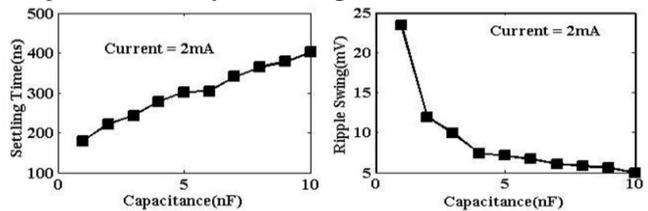

*Fig: 4 Variation of settling Time and Ripple Voltage with*

*3)Variation with Vreference Gap*:
For a constant current(2mA) and fixed number of comparators(8),the settling time as well as the ripple is found to increase with the gap between the successive Vreferences.This is because with increasing Vref-Gap,the effective range over which the LDO control increases, accounting for larger Vripple.Also for larger Vref gap, the change in the state of the driver PMOS will occur after larger time. Hence the settling time also increases.

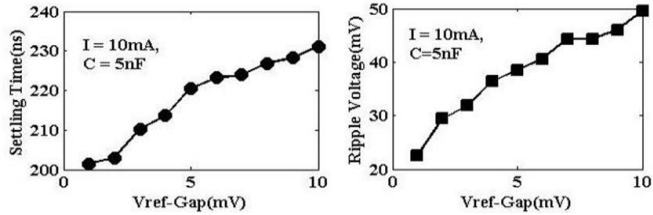

Fig: 5 Variation of settling Time and Ripple Voltage with Vref-Gap

4) *Variation with Minimum Step Size* :

For fixed number of comparators, the Block Size determines the Minimum Step size (i.e., Minimum Step Size = Block Size/Num-C). If the Minimum Step Size(or Block Size) is large, more PMOS can be turned on/off in response to the change. Hence increasing the Minimum Step Size will decrease the settling time and increase the voltage ripple accordingly.

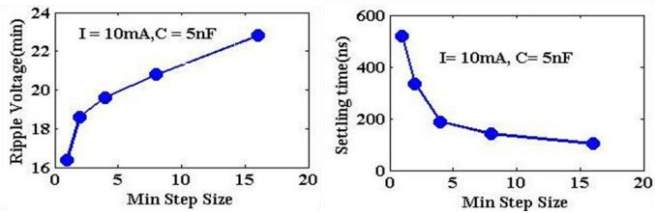

Fig: 6 Variation of settling Time and Ripple Voltage with Min Step

5) *Variation with the Width of the Device*:

Current supplied to the load is mainly determined by the maximum number of transistors that can turn on in one clock cycle. Thus along with the size of the individual transistors, it becomes important to regulate the number of transistors that can be turned ON. As predicted and observed an optimum point is obtained on either side of which the ripple voltage increases with the change in the width of the device. This is because for constant current load and capacitance, there is an optimum value of supply current that needs to be provided so that the voltage ripple remains minimum.

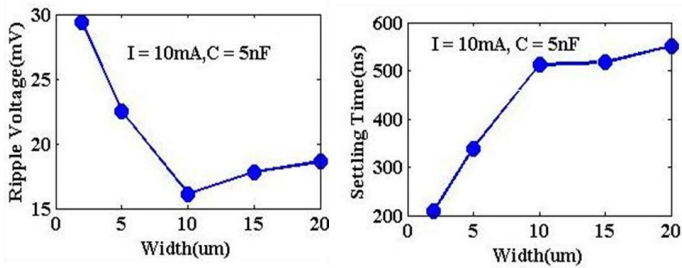

Fig: 7 Variation of settling Time and Ripple Voltage with Vref-Gap

III. TIME INTERLEAVED COMPENSATOR BASED LDO:

A. *DLDO Circuit with Time-Interleaved Comparators:*

The proposed model (fig. 8) comprises of a bank of dynamic comparators clocked in a time interleaved manner. Each dynamic comparator is connected with a PMOS switch. The clock time period is divided into equal phases as the number of comparators. Each comparator is clocked at the beginning of each division through a pulse generation circuit. Each comparator compares the voltage at the output with the required reference voltage at the beginning of each time division and accordingly turns ON or OFF the corresponding PMOS switch.

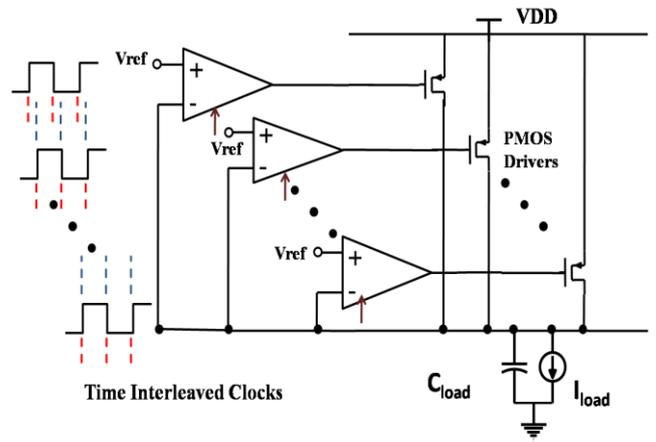

Fig 8 :Schematic Diagram Of Time interleaved comparator based LDO

The working principle resembles the general negative feedback operation. The output voltage is fed back to the comparator. As soon as the comparator senses output voltage is less than the reference voltage, at the next positive clock edge it turns on the corresponding PMOS switch which acts like a current source and supplies more current to the load. Thus the output voltage increases and stabilizes around reference value. All the comparators have only a Single $V_{REF}$.

As all the comparators operate within a single clock period and update the supply current instantly at each of the *N* phases,

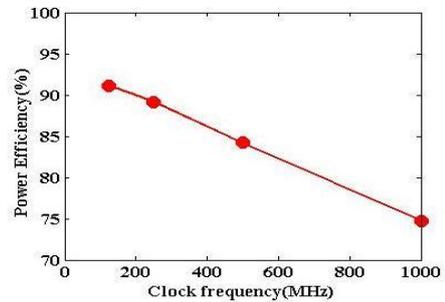

Fig 9 :Variation of Power Efficiency with Clock frequency

(where *N*= number of comparator times within one clock period) the settling time is observed to be significantly small as compared to the single comparator case. On a CMP, the local load current can vary over a large range. The main problem with such configuration is that the efficiency reduces drastically at lower values of current because relatively larger amount of power is burnt in the control circuitry constituting of multiple comparators. For a given load-current, power efficiency of the LDO can be improved by lowering the clock frequency (fig. 9), provided an acceptable level of output ripple is maintained. This explains a need for the optimization of the circuit for minimization of power for low load current. One of the effective ways to achieve this is to dynamically dynamic modify the clock frequency for depending upon load current.

B. *Variation of design parameters:*

The variations of output voltage properties with design parameters are discussed below:

1. *Variation with Clock frequency*:

For a constant load (I = 10.0 mA, C= 3.0 nF) and fixed number of comparators (16), steady state voltage ripple and settling time is found to decrease with increase in clock frequency. As with increase in clock frequency the duration of each time interleave decreases more number of comparisons can be done within fixed amount of time. So for a sufficiently large current step the number of comparisons can be finished quickly amounting to a decrease in settling time. The charging / discharging time also decreases with the increase in clock frequency resulting to a decreased voltage ripple.

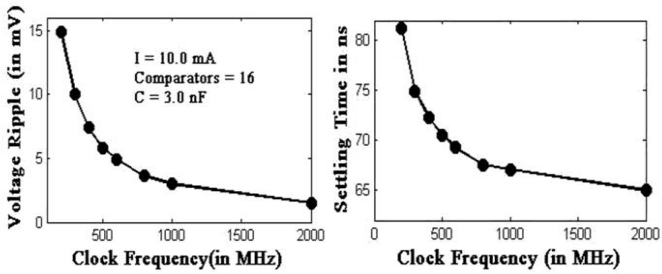

*Fig 10: Variation of ripple voltage and settling time with clock frequency*

2. *Variation with Number of comparators*:

For a constant load current (10.0mA, C= 3.0 nF), fixed total current supplied by the pMOS switches (48 mA) and fixed clock frequency (1GHz), the steady state voltage ripple is found to decrease with the number of comparators (pMOS width= 20um, ). With the increase in comparison levels the duration of the time interleaves decreases, reducing the charging /discharging time of the capacitance and in turn the steady state voltage ripple. However, the settling time required is found to be almost constant because of the balancing effect of number of pMOS switches and current supplied by each of them.

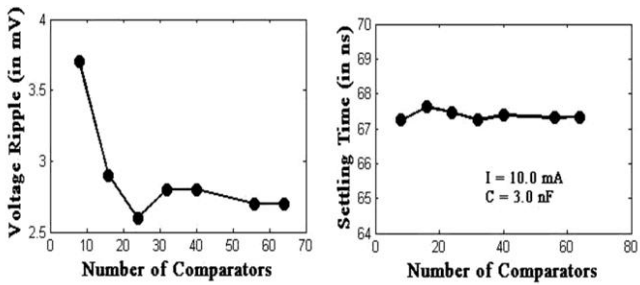

*Fig 11: Variation of ripple voltage and settling time with No of comparators*

3. *Variation with Load current:*

For a constant load current(10.0mA), the settling time is found to increase with Capacitance (Number of compartors = 16, pmOS width= 20um, Clock frequency = 1Ghz). This can be explained by the fact that for constant supply of current, the charging/discharging time of the circuit is directly proportional to the capacitance.Larger capacitance takes more number of time steps to charge from zero level to steady state voltage level. As it takes larger time for larger capacitances to charge/discharge,the ripple for higher value of capacitances is lower(as shown in Fig.L).

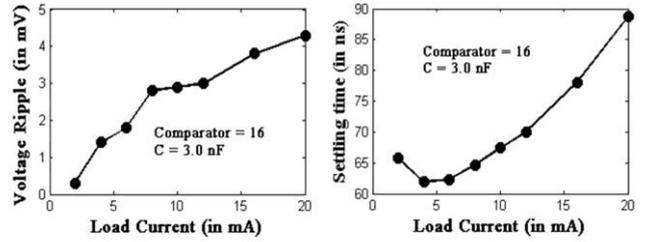

*Fig 12: Variation of ripple voltage and settling time with Load current*

4. *Variation with Output Capacitance :*

For a constant load current(10.0mA), the settling time is found to increase with Capacitance (Number of compartors = 16, pmOS width= 20um, Clock frequency = 1Ghz). This can be explained by the fact that for constant supply of current, the charging/discharging time of the circuit is directly proportional to the capacitance.Larger capacitance takes more number of time steps to charge from zero level to steady state voltage level. As it takes larger time for larger capacitance to charge/discharge,the ripple for higher value of capacitance is lower(as shown in Fig. M).

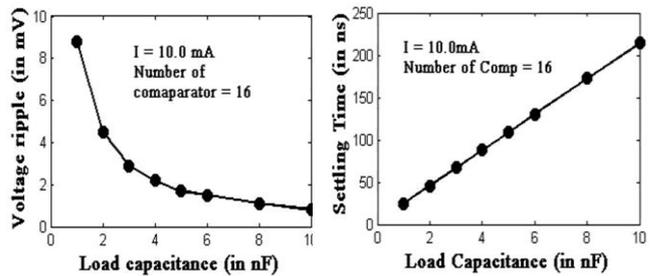

*Fig 13: Variation of ripple voltage and settling time with Load Capacitance*

5. *Variation with width of each PMOS:*

The load current range of the LDO depends on the maximum current that can be pumped through the PMOS switches. Thus along with the maximum number of the individual transistors, it becomes important to regulate the width of transistors that can be turned on.

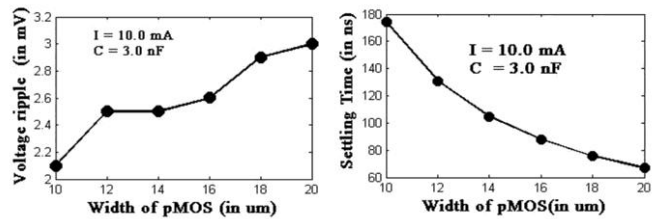

*Fig 14: Variation of ripple voltage and settling time with width of pMOS switches*

The settling time is found to decrease with the width of the device because of the increased ability of the pMOS switch array to pump current in every clock cycle. More current charges the capacitance in less number of clock cycles and decreases the settling time.
For an increase in width pMOS switches pump more current to the load capacitance in each time interleave, this leads to increase steady state ripple. Simulation results also follow the trend for a load current of 10.0 mA, load capacitance of 3.0nF and an array of 16 comparators.

*C. Frequecy Scaling with load current variation:*

Dynamic frequency scaling is achieved through a programmable clock divider whose division factor varies with the local current level. In the DLDO, the number of driver *ON* PMOS transistors at any instant is the indicator of the local current level. The logic comprising of an adder-block, sums all the gate voltages of the PMOS transistors (0 indicates it is on and VDD indicates it is off). (fig. 16), thereby producing the digital control signal for the programmable divider. The proposed circuit has been modeled in Matlab using behavioral equations governing the operation of the DLDO. The rigorous numerical analysis done for design optimization has been corroborated with SPICE simulations for desired operating conditions (fig. 15). The plots show that for lower currents, the clock speed can be proportionately reduced without significantly compromising the ripple while reducing the power in the comparators. Table-I summarizes the results for output voltage ripple for different load current values and corresponding clock frequencies. Nearly constant peak to peak ripple and current efficiency is observed for different load current values in the proposed scheme.

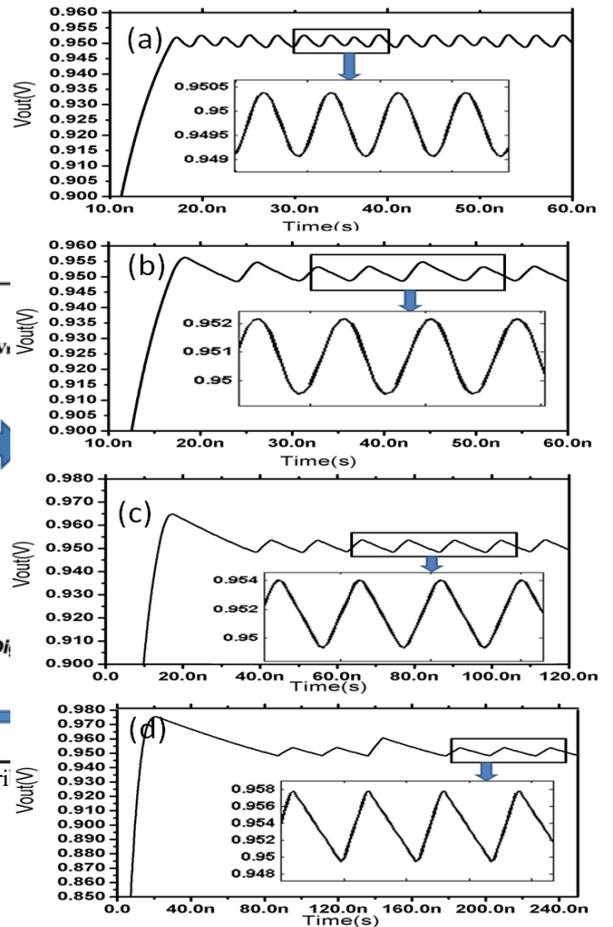

Fig 15: (a) I= 40mA, Clock period = 1ns, (b) I= 20mA, Clock period = 2ns, (c) I= 10mA, Clock period = 4ns, d: I= 5mA, Clock period = 8ns, (Left hand side figures correspond to Matlab Simulations and right hand side figures correspond to Cadence Simulations)

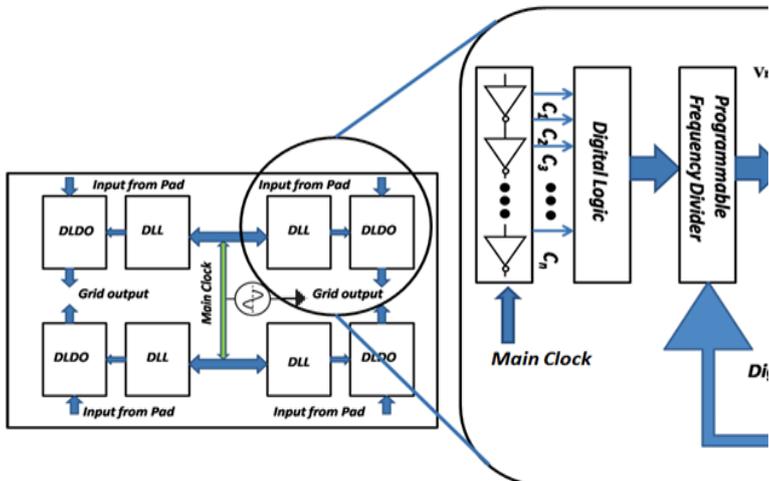

Fig 16: Schematic Diagram of the LDO Scheme with its distri

| | | | |
|---|---|---|---|
| 20 mA | 2ns | 4.4 mV | 97.3% |
| 10 mA | 4ns | 5.0 mV | 96.9% |
| 5 mA | 8ns | 5.2 mV | 96.8% |

## IV. GRID MODELLING

Different IR drop models for flip-chip and wire bond packages have been proposed by [12]- [15]. Modelling of IR drp in a typical flip chip package is done in [16]. Here we adopted the flip chip package for grid modeling (fig. 17).

The most common way to distribute power in a GSI chip is to distribute it through an on-chip grid made of orthogonal segments([17]-[19]). The horizontal and vertical segments of a grid are routed at different metal levels and are connected through vias at the crossing points. The main challenge in the design of the power distribution network is to achieve a minimum acceptable voltage fluctuation across the with minimum routing area of the interconnect metal layers ([20]-[22]).

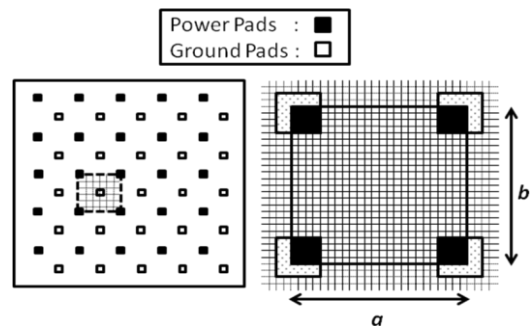

Fig 17: Power and Ground pads for flip chip package (left) and grid between neighbouring four pads (right)

TABLE II: PARAMETERS FOR GRID-MODELLING

| Parameters | Values | Parameters | Values |
|---|---|---|---|
| Cell Size(a) | 1 mm | Resistance (*distributed over 0.1mm*) | 0.55 Ω |
| Max Current Density | 5 A/cm$^2$ | | |
| Segment Length ($l_{segx}$) | 0.1 mm | Capacitance ( *lumped at each LDO node*) | 9 nF |
| Segment Width ($l_{segy}$) | 0.1 mm | | |

### A. Pad-Modelling:

Almost two-thirds of the total pads are used for power distribution. These power and ground pads are uniformly spread throughout the surface of the chip to reduce voltage drop. The pads have been modeled as voltage sources with associated resistances, capacitances and inductances with values as specified in Table III (fig. 18).

TABLE III: PARAMETERS FOR PAD-MODELLING

| Parameters | Values | Parameters | Values |
|---|---|---|---|
| Pad Resistance | 1 Ω | Pad Capacitance | 5pF |
| Pad Inductance | 1 nH | Supply Voltage | 1 V |

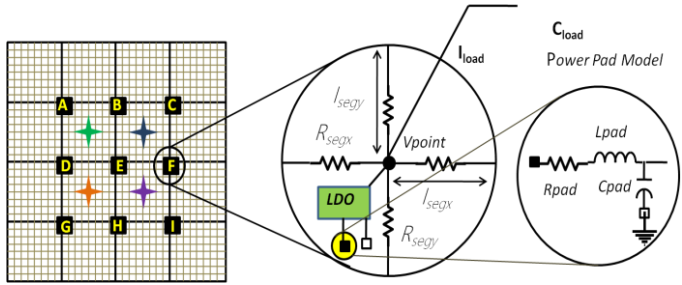

Fig 18: LDO and pads integrated with On-Chip Power Grids. Black points correspond to pad and LDO locations and Colored points correspond to Output points

### B. Positioning of LDOs:

As shown in fig. 18, every LDO is connected to a pad i.e. a LDO covers an effective area of 1mm$^2$. Since the maximum current density simulated is 5 A/cm$^2$, the maximum current supplied by the LDO is 35 mA. The VDD of the LDO is connected to the power supply pad and the node is connected to the points in the grids. An accurate reference voltage is internally generated within the LDO. The load is modeled by an ideal pulsating current source with an effective shunt capacitance of 9nF.

The operation of the LDO was evaluated by providing asynchronous pulse current having different levels and time period distributed randomly over all the 9 pad points. In such a pulsating condition, the voltage variation throughout the grid provided a strong measure of the regulating ability of the proposed circuit scheme.

### C. Parallel Operation of mutliple LDOs

The allocation of the clock frequency to different ranges of load current was found to be an important design parameter. Fig. 19 dpeicts the variation of the output ripple with the choice of frequency allocation, for a load current of 20 mA pr LDO. (Grid Size = 1mm$^2$ )

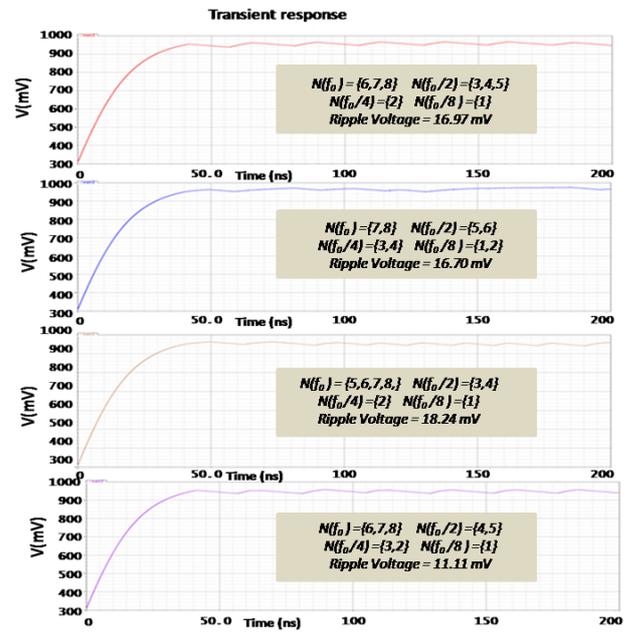

Fig 19: Output of the Cases discussed above(for 4 different configurations) For Load Current (20 mA) for all the cases, N{f} means working frequency is f when number of *ON* PMOS is either of the elements in the right hand side brackets. N{fo/2} = {3,4} implies choosing clock frequency = 1GHz/2 for 3 or 4 PMOS being *ON*.

The plots show significant dependence of frequency band allocation to different load current levels. Hence, this distribution needs to be optimized based on simulation based analysis. The ripple was also fund to improve for a smoother and uniform frequency gradient with respect to load current.

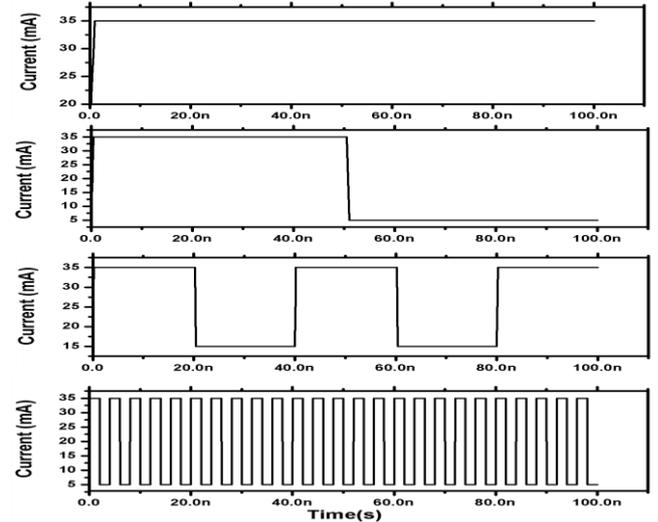

Fig 20 : Skewed Load current pulses(from top-wise current at locations A to I only four are shown in the figure)

In SPICE simulation, the grid-performance has been simulated over an effective area of 16 mm$^2$ with 9 LDOs present at the position of the Power Supply Pads as shown in Fig 6. Also the distribution of current waveforms in the various sections is presented in Figure 9. All the nodes in the grid behave very closely even in very random distribution of local current. The grid-level simulation results show the advantage of the proposed scheme, in terms of fast response with stable, low peak to peak ripple for a wide range of load currents.

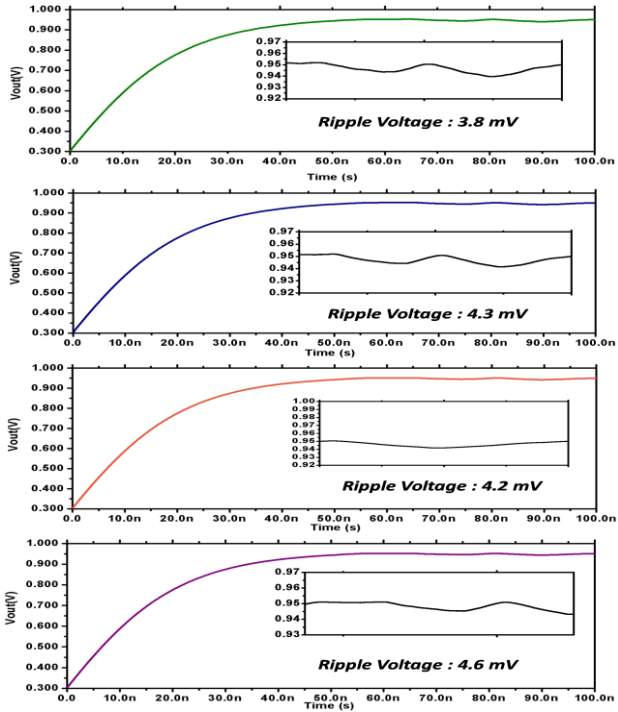

Fig 21: Cadence Simulation Results for Vref = 900 mV(Individual Levels – top) The colors correspond to the marked points(in fig F) (Ripple is shown in offset)

Results of the proposed design compare very favorably with results reported earlier [23]. The proposed design achieves orders of magnitude faster transient response for a much larger load current, while achieving similar current efficiency and output ripple. For the maximum load current of 20mA, the effective clock frequency increase as compared to the work in [23] is 8x(1GHz/10MHz) = 800, which is the main factor conducive in achieving low ripple along with faster response time at comparable efficiency.

TABLE IV
Performance Comparison

|  | [23] | This work |
|---|---|---|
| Technology | 65nm | 65nm |
| Load current | 200μA | 2 to 20mA |
| Current efficiency | 98.7% | 96.8% |
| Output ripple | 3mV | 5mV |
| Clock freq. | 10MHz | 1GHz to 100MHz |
| VDD | 0.5V | 0.7V |
| Response time | 240μs | ~100ns (interleave factor =8) |

## V. CONLUSION

We proposed the design of digital LDOs with multiple, time-interleaved dynamic comparators that can provide low ripple and fast response time. We employed load dependant clocking frequency to reduce the power overhead due to larger number of comparators and hence avoided degradation in current efficiency over a single comparator design. Simulation based analysis shows that the DLDO with time-interleaved comparators can achieve better overall performance in terms of current efficiency, ripple and settling time.